\documentstyle[preprint,aps,prl]{revtex}

\input epsf
\begin{document}
\draft

\title{How to make a tiny black hole?}
\author{Piotr Bizo\'n}
\address{Department of Physics, Jagiellonian University, Cracow, Poland\\
bizon@thp.if.uj.edu.pl} 

\date{\today}
\maketitle

\begin{abstract}
This is a brief review of critical phenomena in gravitational collapse.
The conceptual issues are emphasized and some directions for future  research
are suggested. The paper is not addressed to the experts in the field -- for
them  little will be new. It is rather meant to introduce others
into one of the most rapidly developing areas of research in general
  relativity with the hope of attracting them into the subject.
\end{abstract}
\vskip2pc

The gravitational collapse of matter leading to a formation of a black
hole is probably the most fascinating prediction of general relativity.
On the one hand, this phenomenon has significant astrophysical consequences
since it is believed to be the end-point of evolution of massive stars.
On the other hand, the understanding of the dynamics of gravitational
 collapse is a major theoretical challenge in general relativity
and a subject of intensive studies.
One of the main goals of these studies is to
confirm or falsify the cosmic censorship hypotesis,
which, loosely speaking, says that a physically realistic gravitational
collapse cannot result in a naked singularity, i.e. a singularity
visible to a distant observer.
Although it would be most interesting to assert the validity of the
cosmic censorship hypothesis for the vacuum Einstein equations, this
problem is too hard because, at present,  only the spherically symmetric 
equations are tractable by analytical techniques, and, unfortunately,
in this case the vacuum Einstein equations become trivial. Thus, to
have nontrivial spherically symmetric dynamics it is indispensable to
couple matter fields. 

From this theoretical perspective, the physical relevance of
a matter model is of secondary importance -- the priorities are simplicity
 and generality.
The simplest choice is a minimally coupled linear massless 
scalar field -- this system reduced to spherical symmetry
was studied by Christodoulou in a series of papers (see\cite{c1,c2} and 
references
therein).
Analyzing
the evolution of asymptotically flat regular initial data, he
showed that "weak" (in the sense of a certain function norm physically
corresponding to a measure of energy concentration) initial data  posses global
time evolution which asymptotes the flat spacetime, whereas "strong" initial
data collapse to a black hole. These results suggested that for a
 one-parameter
family of initial data interpolating between "weak" and "strong" regime,
there would be a critical parameter value corresponding to the treshold of
black hole formation.
 If so, there arises a natural question: what is the mass of the
black hole at the treshold? Is it 
infinitesimal, or is there a finite mass gap in the 
spectrum
of black hole masses? I will refer
to these two possibilities as to type II and type I behaviours, 
respectively\footnote{This terminology originates from the formal analogy 
with first
and second order phase transitions. Notice that type I
behaviour is typical in the astrophysical context with the mass gap being of
the order of the Chandrasekhar mass for fermionic matter.}.

The program of studying this question numerically was initiated and carried
out with remarkable success by Choptuik\cite{matt}. His results can be
 summarized
as follows.
Consider a one-parameter interpolating
family of initial data for the spherically symmetric 
Einstein-massless scalar field equations. A typical initial
 profile for the scalar field is
an ingoing  Gaussian wave with adjustable amplitude. With such a profile
one can associate two length scales: the physical extent (the width), $L$,
and the Schwarzschild radius $R_S$. 
The dimensionless parameter, $p=R_S/L$, which is monotonically related to
the amplitude of the Gaussian, characterizes the degree
of  energy concentration. In agreement with Christodoulou's results,
for weakly coupled data, $L>> R_S$,
the ingoing wave implodes through the center and then escapes to infinity
leaving behind the flat spacetime. When $L \approx R_S$, the gravitational
interaction becomes important, the energy of the imploding wave is partially
trapped and a black hole forms.
 Let us denote by $p^{*}$ a critical value of
the parameter which separates black-hole-spacetimes ($p>p^{*}$) from
no-black-hole ones ($p<p^{*}$). Given two values $p_{\rm weak}$ and
$p_{\rm strong}$, it is (in principle) straightforward to find $p^{*}$ 
by bisection.

 This problem looks so natural that one might wonder
why it had not been studied earlier. Actually it had, but the results were
inconclusive because of insufficient numerical accuracy.
 The point is that the features of near-critical solutions are
exponentially sensitive to the distance from the treshold $|p-p^{*}|$
and, as this distance tends to zero, there appears an oscillating structure
on progressively smaller spatio-temporal scales. To probe this structure
 it was instrumental to use a sophisticated
numerical code. Choptuik used an adaptive
 mesh-refinement algorithm which
allowed him to approach the treshold almost down to the machine precision
 $|p-p^{*}|/p^{*} \approx 10^{-15}$. The effort Choptuik invested in 
implementing
his algorithm payed with interest -- having a high resolution code, he was 
not 
only able to resolve the mass-gap
question, but also, as a premium, found unexpected and intriguing phenomena
at the treshold of black hole formation.

The most important observation was the {\em universality} of critical 
behaviour. 
Here universality refers to the fact that, in the so called intermediate
asymptotics (i.e. before a solution "decides"
whether to collapse or to disperse), all
near-critical solutions have the {\it same} (i.e. family-independent) shape
in the strong-field region (i.e. near the center of implosion). 
In other words, in this asymptotic regime the details of initial data are 
washed out.
The precisely critical solution ($p=p^{*}$), called a choptuon,
  has an unusual symmetry of {\em discrete self-similarity}, that is it 
reproduces itself (echoes) on progressively finer scales:
$r \rightarrow r e^{-\Delta}$, $t^{*}-t \rightarrow (t^{*}-t) e^{-\Delta}$,
where $t^{*}$ is the accumulation time
of successive echoes and a constant $\Delta \approx 3.44$.

The second important result was the resolution of the mass gap problem.
Choptuik has found that near-{\em super}critical data form black holes with
masses satisfying the power-law
\begin{equation}
M_{BH} \simeq C (p-p^{*})^{\gamma} , 
\end{equation}
where the proportionality constant $C$ is family-dependent, but the critical
 exponent $\gamma \approx 0.37$ is again universal. Thus, by fine-tuning the 
parameter $p$, one can
make a black-hole of arbitrarily small mass\footnote{Perhaps it is worth
stressing that the absence of a mass 
gap
for black holes in
  the 
Einstein-massless scalar field system, is {\em not}
a trivial consequence of the scale invariance of the model -- the latter
 implies only that a mass gap
 could not be universal.}. This gives the answer to the
question posed in the title. At this point
I should warn the reader, who might be anxious to produce a tiny 
black hole in this manner in his lab, that the experiment is 
dangerous -- the smaller the black hole,
the stronger the gravitational field at the horizon!

The choptuon is a limiting, sort of zero-mass black hole.
In physical terms, it can be viewed
 as a collapsing radiating ball of field energy for which the rate of
collapse is exactly balanced by the rate of energy loss by radiation,
so that when the ball shrinks to zero radius all of its energy is radiated
away\cite{he1}.
Due to the accumulation of echoes, the curvature of the critical solution 
diverges at the origin as $t \rightarrow t^{*}$. This singularity is 
visible from
null infinity, as Hamad\'e and Stewart have demonstrated\cite{hs}, so, strictly
 speaking, 
the choptuon constitutes a counterexample to the cosmic censorship
hypothesis. 
On the one hand, this counterexample is disturbing because it shows that
the evolution of perfectly regular initial data for the realistic matter
may lead to the formation of regions
 of spacetime with arbitrarily large curvature which are
 not
surrounded by a event horizon. On the other hand, the phenomenon is not
 generic --
the naked singularity can be destroyed by an arbitrarily small perturbation.

Soon after Choptuik's discovery a similar critical behaviour  has 
been observed
in several other models of gravitational collapse with different 
sources\cite{ec,hhs}
and even, in one notable case, without spherical symmetry\cite{ae}. 
I shall not go into the details of these models -- let me only note that
the overall picture of criticality is qualitatively the same as in the scalar
field collapse, possibly with one difference: in certain models the critical 
solution
is not discretely, but {\em continuously} self-similar. These studies
 have lent support to the conjecture
 that
such features  like 
universality, black-hole mass scaling, and self-similarity (discrete or
continuous) are the robust properties of type II gravitational
collapse\footnote{Numerical values of a critical exponent $\gamma$
and a periodicity scale $\Delta$ {\em do} depend on a model. Actually,
in the first three analyzed models (scalar field, axisymmetric gravity,
radiation fluid) the critical exponent $\gamma$ had the same value (up
to numerical errors), which suggested the universality in the 
broader sense of model-independence. More accurate recent 
computations strongly suggest that this fact was just 
a misleading numerical coincidence.}.

Having emphasized the genericity of type II critical collapse, it is
appropriate to point out that the above mentioned models
share a common, rather restrictive, property: they do not have regular
stationary solutions. This fact reduces the long-time outcomes of 
evolution basically to two possibilities: collapse or 
dispersal\footnote{A priori there is also a possibility of chaotic
evolution, however, to my knowledge, it has not been observed in this
context.}. Clearly, in a model 
having a {\em stable} stationary
solution, there is an additional possibility that this solution will be 
the end-point of evolution of some (presumably large)
set of initial configurations -- after all this is how the stars have been
formed. An even more interesting situation arises when a 
model admits an 
{\em unstable} stationary solution with exactly one unstable mode. Such a 
solution 
can play a role of a critical configuration separating collapse from
dispersion, thereby giving rise to type I critical behaviour.
This phenomenon has been found very recently in the Einstein-Yang-Mills
model\cite{eym}, where depending on a class of initial data, both
 types, I and II, 
of critical behaviour are present. An interesting implication of that
is the existence of a critical line in the parameter space which interpolates
between theses two types. The transition point lying on this line,
corresponding to a sort of superposition of both critical solutions,
can be located by fine-tuning  the parameters of
certain {\em two}-parameter families of initial  data. 
Thus, the presence of stationary solutions in a model
enriches the possible long-time asymptotics and may lead to  
complex phase diagrams. I am confident
that the analysis of\cite{eym}, which has opened this issue, will be followed
 by further  activity on
related models.

The analytical understanding of the numerical phenomenology of critical
 behaviour
described above is a great theoretical challenge. Although at present there
are no rigorous results in this area, a substantial progress has been made
on a heuristic level. First of all, we have a fairly convincing picture of
the origin of universality. 
The main assumption of the mechanism explaining universality is that
the critical solution has exactly one unstable mode\cite{koike}. Then the 
stable
manifold
$W_S$ of the critical solution has codimension one and separates (at least
locally) the phase space of a model into spacetimes contaning a black hole and
spacetimes that do not. An interpolating one-parameter family of initial data
is a curve in the phase space which intersects the stable manifold $W_S$
at the critical parameter value $p=p^{*}$. The critical data are attracted
 along $W_S$
towards the critical solution (for this reason the critical solution is
sometimes called a codimension one attractor). The near-critical data, by
 continuity, 
initially remain close
to $W_S$ and approach the critical solution (intermediate asymptotics), but
ultimately are repelled from $W_S$ by the  growing unstable mode. 
\vskip1pc
\centerline{\epsfbox{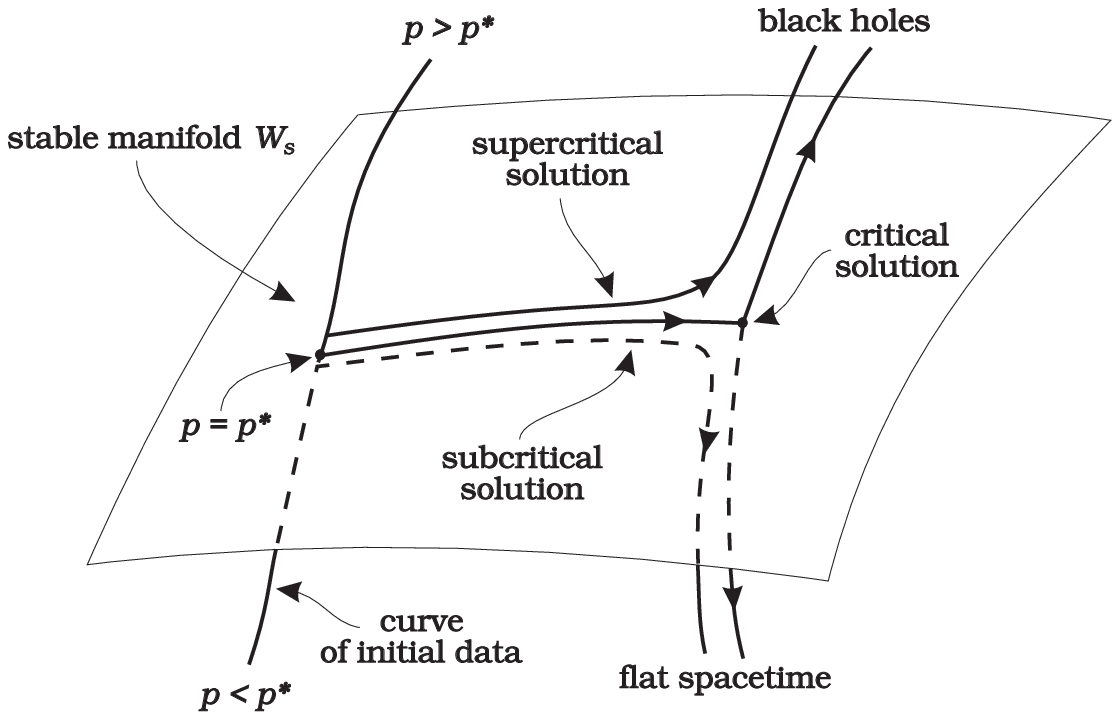}}
\centerline{A sketch of the near-critical evolution.}

The fact that the {\em same} unstable mode dominates the long-time behaviour
of {\em all} near-critical initial data is the origin of 
universality.
 It should be stressed that this mechanism explains the critical
 dynamics controlled by
a {\em  given} codimension one attractor. If there are multiple attractors in
 the
 phase space, there arise crossover phenomena associated with transitions
between different basins of attraction. For example, 
this happens in 
 the Einstein-Yang-Mills model mentioned above, where the transition point
representing the coexistence of two types of critical behaviour is nothing
else but the
  codimension {\em two} intersection of the stable manifolds of two critical
solutions.

This framework gives also a natural explanation of the power-law (1),
and, more importantly, provides a method of computing the critical exponent
 $\gamma$\cite{ec,koike}. For continuously self-similar 
critical solutions the argument goes as follows. As I wrote above, the 
evolution of a small
perturbation around the critical solution is dominated by the single
unstable mode. The amplitude of this mode is proportional to 
$|p-p^{*}| (t^{*}-t)^{-\lambda}$,
where the Lyapunov exponent $\lambda$ is positive.
 Assuming that this perturbation leads to collapse ($p>p^{*}$), it follows
 from 
dimensional analysis that 
the scale of mass of a resulting black hole is set by the time in which
the perturbation grows to a finite size: 
$(p-p^{*}) (t^{*}-t)^{-\lambda} \sim O(1)$.
Hence,
$M_{BH} \sim (p-p^{*})^{1/\lambda}$, and the comparison with (1) 
yields
$\gamma=1/\lambda$. Thus, to compute $\gamma$ it suffices to find a critical 
solution\footnote{A critical solution can be computed by inserting 
the self-similarity ansatz into the field equations. In the case of
continuous self-similarity
this reduces the problem to solving a system of ODE's. The case of
discrete self-similarity is more complicated -- here one obtains a
sort of nonlinear hyperbolic eigenvalue problem for the periodicity
scale $\Delta$\cite{g1,g2}.}
and then, using linear stablity analysis, calculate the
Lyapunov exponent of the unstable mode. The critical exponents have
been computed in this manner in several models\cite{m,he1,hhs}. 
The fact that these calculations have  reproduced the  values   obtained
 from dynamical simulations, confirms the correctness
of the overall picture. In the case of a
discretely self-similar critical solution the calculation is technically
much more involved but the basic idea remains the same\cite{g2}.

The mechanism described above is analogous to the standard renormalization 
group (RG) description
of  second order phase transitions in statistical physics.
 Actually, it is
more than analogy -- the time evolution can be viewed as the flow
in the phase space generated by the iteration of  
RG transformation\cite{gold,bk} (which amounts to a suitable 
rescaling of variables). 
The critical solution corresponds to a fixed point
 (continuous self-similarity) 
or a limit cycle (discrete self-similarity) of the RG transformation\cite{a}.
This approach is very useful in determining the so called {\em irrelevant}
terms in the evolution equations. For example, it has been observed by
Choptuik that the inclusion of a mass or self-interaction term to the 
Einstein-massless scalar field equations 
does not affect {\em quantitatively} the critical behaviour. In other words,
all  models of the  minimally coupled gravitating scalar field with different
self-interaction potentials belong to the same universality class (here 
universality refers 
to model-independence). This follows
immediately from the RG transformation for the scalar field which 
 has the form
\begin{equation}
\phi_L(r,t^{*}-t) \equiv R_L \phi(r,t^{*}-t) = 
\phi(r/L,(t^{*}-t)/L) \, , 
\qquad L>1. 
\end{equation}
It is easy to check that the "potential" terms in the field equations
for $\phi_L$ are multiplied by $L^{-2}$ (as compared to the "kinetic" terms)
and therefore after many iterations these terms become negligible, ergo
 their presence does not change the 
long-time asymptotics.

Although the ideas borrowed from the theory of phase transitions
in statistical physics have been  heuristically very helpful in 
understanding the critical behaviour in gravitational collapse, 
the speculations about possible deep {\em physical} connection between these 
phenomena are not,
in my opinion, justified  and belong rather to an art of writing
introductions in the Physical Review Letters. 
This remark applies in particular to the interpretation of the black hole
mass as an order parameter, which is based on the striking similarity of the
power-law (1) with the analogous formula for the spontanous magnetisation
in the ferromagnetic/paramagnetic phase transition. 
It seems to me that instead of taking such analogies too literally, it would
be more fruitful to search insight in toy-models exhibiting dynamical
phase transitions. Let me illustrate what I mean with the following example.

 Consider a bistable dissipative gradient system
\begin{equation}
\frac{\partial f}{\partial t} = - \nabla E(f) \, .
\end{equation}
Here the energy functional $E(f)$, whose gradient flow generates the 
evolution, is
assumed to  be bounded from below and  to have exactly two 
minima $f_{\pm}$. If, in addition, the space of finite energy functions is
connected, then one can show (modulo technicalities\footnote{For the sake 
of brevity I am 
 cavalier
about the mathematical "details" (such as an important issue of
 compactness). I simply assume  tacitly that
the functional $E(f)$, when defined in a suitable function space, has all
 needed
 properties.}) by minimax argument that there exists a
 saddle
point of energy (denote it by $f^{*}$). These three extrema of energy are
the stationary solutions of eq.(3), two stable ones and one unstable.
It follows from (3) that the energy decreases in time, so it is natural
to expect (and in fact can be proven in many cases) that these stationary 
solutions are the only possible end-points
of the evolution (assuming that global evolution exists). Of course, generic
initial data will flow to one of the stable solutions $f_{\pm}$. By definition
such data comprise respective basins of attraction, $W_S^{\pm}$, of these 
solutions.  A codimension one boundary between $W_S^{-}$ and $W_S^{+}$ 
 is a stable manifold, $W_S^{*}$, of the
saddle point $f^{*}$. Now, consider, as above, a one-parameter family $f_p$ of
 initial data interpolating between $W_S^{-}$ and $W_S^{+}$. As usual, let
 $p^{*}$ be the critical parameter value  for which $f_p$ intersects
 $W_S^{*}$. In the intermediate asymptotic regime the evolution of 
near-critical 
initial data 
is dominated
by the single unstable mode of $f^{*}$
\begin{equation}
f_p(t) \simeq f^{*} + C (p-p^{*})\, e^{\lambda t}\, \xi \, ,
\end{equation}
where $\xi$ is the eigenmode associated with the positive eigenvalue
$\lambda$. Here the amplitude $C$ is the only vestige of initial data.
 Depending on the sign of $C$, the solution $f_p(t)$ will ultimately evolve
towards $f_{+}$ or $f_{-}$. The "lifetime" $T$ of the near-critical solution
staying in the vicinity of $f^{*}$ is determined by the condition
$|p-p^{*}| e^{\lambda T} \sim O(1)$ which gives
$T \sim (-1/\lambda) \ln|p-p^{*}|$.

This  model reflects quite well certain features of type I critical
collapse observed in the Einstein-Yang-Mills model. Of course there are
substantial differences, but one cannot expect too much from such a simple 
model. It would be most interesting to construct a toy-model
of type II critical behaviour. This could give insight into the origin
of discrete self-similarity, which is widely considered as the most
mysterious feature of critical collapse. Let me end with the speculation that,
in this respect, there might be a remote mathematical connection between
echoing and the dynamical formation of fine structure in certain material
phase transformations, where the flow of
energy to higher and higher wavenumbers can be understood in terms of models
 described by eq.(3) with the energy functional which does not attain a minimum
 but has minimizing sequences with finer and finer structure\cite{ball}.

{\em Acknowledgement.} This research was supported in part by the KBN grant
 2/PO3B/091/11.

\end{document}